\documentclass[aps,prc,groupedaddress,showpacs,manuscript]{revtex4}

\usepackage{amsmath}
\usepackage{graphicx}
\usepackage{colordvi}
\usepackage{ulem}

\begin{document}

\title{UrQMD calculations of two-pion HBT correlations in central Pb-Pb collisions at $\sqrt{s_{NN}}=2.76$ TeV }

\author {Qingfeng Li$\, ^{1,2,4}$\footnote{E-mail address:
liqf@hutc.zj.cn}, G. Gr\"af$\, ^{2,3}$ and Marcus Bleicher$\,
^{2,3}$}
\address{
1) School of Science, Huzhou Teachers College, Huzhou 313000,
P.\ R.\ China \\
2) Frankfurt Institute for Advanced Studies (FIAS), Johann Wolfgang Goethe-Universit\"{a}t, Ruth-Moufang-Str.\ 1, D-60438 Frankfurt am Main, Germany\\
3) Institut f\"{u}r Theoretische Physik, Johann Wolfgang Goethe-Universit\"{a}t, Max-von-Laue-Str.\ 1, D-60438 Frankfurt am Main, Germany\\
4) Institute of Theoretical Physics, Chinese Academy of Sciences, Beijing 100080, P.\ R.\ China\\
 }


\begin{abstract}
Two-pion Hanbury-Brown-Twiss (HBT) correlations for central Pb-Pb
collisions at the Large-Hadron-Collider (LHC) energy of
$\sqrt{s_{NN}}=2.76$ TeV are investigated for the first time with
the microscopic transport model UrQMD (Ultra-relativistic Quantum
Molecular Dynamics). The transverse momentum dependence of the
Pratt-Bertsch HBT radii is extracted from a three dimensional
Gaussian fit to the correlator in the longitudinal co-moving system
(LCMS). Qualitative agreement  with the ALICE data is obtained,
however $R_{out}$ is overpredicted by nearly 50\%. The LHC results
are also compared to data from the STAR experiment at RHIC. For both
energies we find that the calculated $R_O/R_S$ ratio is always
larger than data, indicating that the emission in the model is less
explosive than observed in the data.
\end{abstract}


 \maketitle

\section{Introduction}
In order to shed light on a large number of unsolved questions in
fundamental physics
\cite{Weiglein:2004hn,Gianotti:2004qs,Baltz:2007kq}, the Large
Hadron Collider (LHC) at CERN had been designed,  installed, tested,
and repaired in the past two decades and finally, started normal
operation in the end of the year 2009. Since then, a tremendous
amount of experimental data in various aspects of high energy
physics has been obtained and received much attention by theoretical
physicists. Although the most exciting prediction, the {\it Higgs}
boson, has not been observed in the data from the LHC experiments
\cite{Brumfiel:2011aa}, the extracted bulk properties of the high
temperature fireball created in such ultra-relativistic collisions
have provided unprecedented information for fundamental
investigations of the phase diagram of Quantum Chromodynamics (QCD).
Here we want to explore the expansion properties of the created
matter by investigating the spatial shape of the fireball. Although
it is known that one can not measure the emission time pattern
and/or the spatial profile of the source directly, a
well-established technique, called ``femtoscopy'' or ``HBT'' (see
e.g. \cite{Lisa:2005dd} and references therein) can be employed to
obtain this information. Femtoscopy has been extensively used in the
heavy ion community since it provides the most direct link to the
lifetime and size of nuclear reactions. The ALICE collaboration has
published first results of two-pion Bose-Einstein correlations  in
both p-p \cite{Aamodt:2011kd} and central Pb-Pb \cite{Aamodt:2011mr}
collisions at LHC energies in the beginning of the year 2011. These
experimental results have attracted the research interest of several
theoretical groups
\cite{Aamodt:2011mr,Karpenko:2011qn,Humanic:2010su,Werner:2011fd},
whose models are based on hydrodynamic/hydrokinetic and hadronic
microscopic approaches.

In this paper we show for the first time results for the HBT radii
of two-pion correlations from central ($<5\%$ of the total cross
section $\sigma_T$) Pb-Pb collisions at the LHC energy
$\sqrt{s_{NN}}=2.76$ TeV from the Ultra-relativistic Quantum
Molecular Dynamics (UrQMD) model
\cite{Bass:1998ca,Bleicher:1999xi,Petersen:2008kb,Petersen:2008dd}.
The calculations are compared to ALICE data as well as to those at
the RHIC energy $\sqrt{s_{NN}}=200$ GeV. The UrQMD calculation
results for p-p collisions at LHC energies are presented in
\cite{Graef:2012za}.

The paper is arranged as follows: In Section 2, a brief description
of the UrQMD model and the treatment of the HBT correlations as well
as the corresponding Gaussian fitting procedure is shown. Section 3
gives the main results of the model calculations. Finally, in
Section 4, a summary is given.

\section{Brief description of the UrQMD model and the HBT Gaussian fitting procedure}

UrQMD \cite{Petersen:2008kb,Petersen:2008dd} is a microscopic many-body approach to p-p, p-A, and
A-A interactions at energies ranging from SIS up to LHC. It is based on the
covariant propagation of mesons and baryons. Furthermore it includes rescattering of particles, the
excitation and fragmentation of color strings, and the formation and decay of hadronic resonances. At
LHC, the inclusion of hard partonic interactions in the initial stage is important and is treated via
the PYTHIA \cite{Sjostrand:2006za} model.

In the present study, the cascade mode of the latest version (v3.3)
of UrQMD is used (for details of version 3.3. see
\cite{Petersen:2008kb,Petersen:2008dd}). Some predictions and
comparison works with data from reactions at LHC have already been
pursued based on this version and showed encouraging results for the
bulk properties \cite{Mitrovski:2008hb,Petersen:2011sb}.

To obtain HBT radii, first, about 200 and 10000 central events are
calculated for Pb-Pb collisions at LHC and for Au-Au at RHIC,
respectively. Then, all particles with their phase space
distributions at their respective freeze-out time (last collisions)
from UrQMD are put into an analyzing program using the formalism of
the well known ``correlation after-burner'' (CRAB)
\cite{Pratt:1994uf}. At last, the constructed two-pion HBT
correlator (regardless of charge) in the longitudinally co-moving
system (LCMS) \cite{Pratt:1986cc,Bertsch:1988db} without influence
of residual interactions is fitted (using the $\chi^2$ method) with
a three-dimensional Gaussian form expressed as

\begin{equation}
C(\mathbf{q},\mathbf{K})=1+\lambda(\mathbf{K})
{\rm exp}\left [- \sum_{i,j=O,S,L} q_i q_j R_{ij}^2(\mathbf{K}) \right ]. \label{fit1}
\end{equation}
In Eq.~(\ref{fit1}), $\lambda$ represents the fraction of correlated
pairs \cite{Lisa:2005dd} and $q_i$ is the pair relative momentum
$\mathbf{q}=\mathbf{p}_1-\mathbf{p}_2$ in the $i$ direction, $p_i$
being the momenta of the particles. $L$ is the longitudinal
direction along the beam axis, $O$ the outward direction along the
transverse component of the average momentum $\mathbf{K}$ of two
particles ($\mathbf{k}_T=|\mathbf{p}_{1T}+\mathbf{p}_{2T}|/2$) and
$S$ the sideward direction perpendicular to the afore mentioned
directions. The effect of cross terms  with $i\neq j$ on the HBT
radii is found to be negligible in the present fits when a
pseudorapidity cut $|\eta|<0.8$ is used, as in experiments, and is
not discussed in this paper.

For central collisions, the HBT radii are, except for an implicit
$\mathbf{K}_T$ dependence, related to regions of homogeneity by
\cite{Wiedemann:1999qn}
\begin{align}
R_O^2 & =  \left < (x-\beta_T t)^2 \right >  =  \left < x^2 \right > - 2 \left < \beta_T t x \right >  + \left <\beta_T^2 t^2 \right > , \label{eqn:RO} \\
R_S^2 & =  \left < y^2 \right > , \label{eqn:RS}\\
R_L^2 & =  \left < (z-\beta_L t)^2 \right >  =  \left < z^2 \right > - 2 \left < \beta_L t z \right > + \left < \beta_L^2 t^2 \right > , \label{eqn:RL}
\end{align}
where $x$, $y$, $z$ and $t$ are the spatio-temporal separation of
the particles in a pair and {\boldmath$\beta$}$ = \mathbf{K}/K_0$.
If no space-momentum correlations are present the regions of
homogeneity and the source size coincide. In central collisions the
relation $\left < x^2 \right > \simeq \left < y^2 \right >$ is
satisfied. Thus $R_O^2$ and $R_S^2$ differ mainly in the last two
terms of Eq.~(\ref{eqn:RO}). The first of these two terms is
dependent on the strength of the correlation of emission time and
transverse emission position, while the second one is especially
sensitive to the particle emission duration.

\section{Results}

\begin{figure}
\includegraphics[angle=0,width=0.8\textwidth]{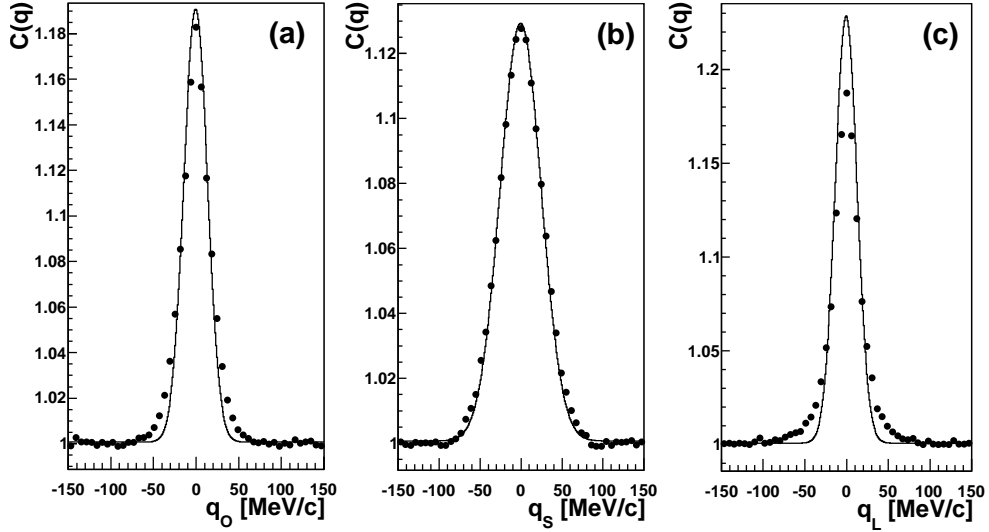}
\caption{ Projections of the three-dimensional correlation function
(points) and of the respective fit (lines) for the $k_T$ bin
$200-300$ MeV/c and $|\eta|<0.8$. When projecting on one axis the
other two components are restricted to the range (-30 $< q <$ 30)
MeV/c.} \label{fig1}
\end{figure}

The correlation functions are studied in bins of the transverse
momentum $k_T=|\mathbf{k}_T|$. Fig.\ \ref{fig1} shows the
projections of the three-dimensional correlation function (points)
and of the respective fit (lines) for the $k_T$ bin $200-300$ MeV/c.
It is seen clearly that the correlator in sideward direction can be
described by a Gaussian form fairly well. However, it deviates
slightly from a Gaussian in the other two directions, especially in
the longitudinal direction, as found and discussed in previous
publications for HICs at lower energies \cite{Li:2008bk}. At LHC,
the fraction of excited unstable particles is much larger than at
lower beam energies, therefore, the non-Gaussian effect is more
pronounced in the current calculations. At RHIC energies, the
non-Gaussian effect was also seen in the experimental HBT
correlator, especially in the longitudinal direction
\cite{Adams:2004yc}. Furthermore, when comparing our fitting result
in Fig.\ \ref{fig1} with that observed in experiment (in Fig. 1 of
Ref.\ \cite{Aamodt:2011mr}), it is seen that the non-Gaussian effect
is stronger in our calculations than in experiment. This might be
due to the omission of all potential interactions between particles
in the current cascade calculations. Support for this interpretation
was found in Ref.\ \cite{Li:2008bk}, where the consideration of a
mean-field potential plus Coulomb potential significantly reduced
the non-Gaussian effect on correlators of pions from HICs at AGS
energies. Therefore, both a dynamic treatment of the particle
transport with a proper equation of state (EoS) for the QGP phase
and the hadron phase, and further theoretical development of the
fitting formalism are equally important for a more precise
extraction of spatio-temporal information of the source
\cite{Pratt:2008bc}.

\begin{figure}
\includegraphics[width=0.495\textwidth]{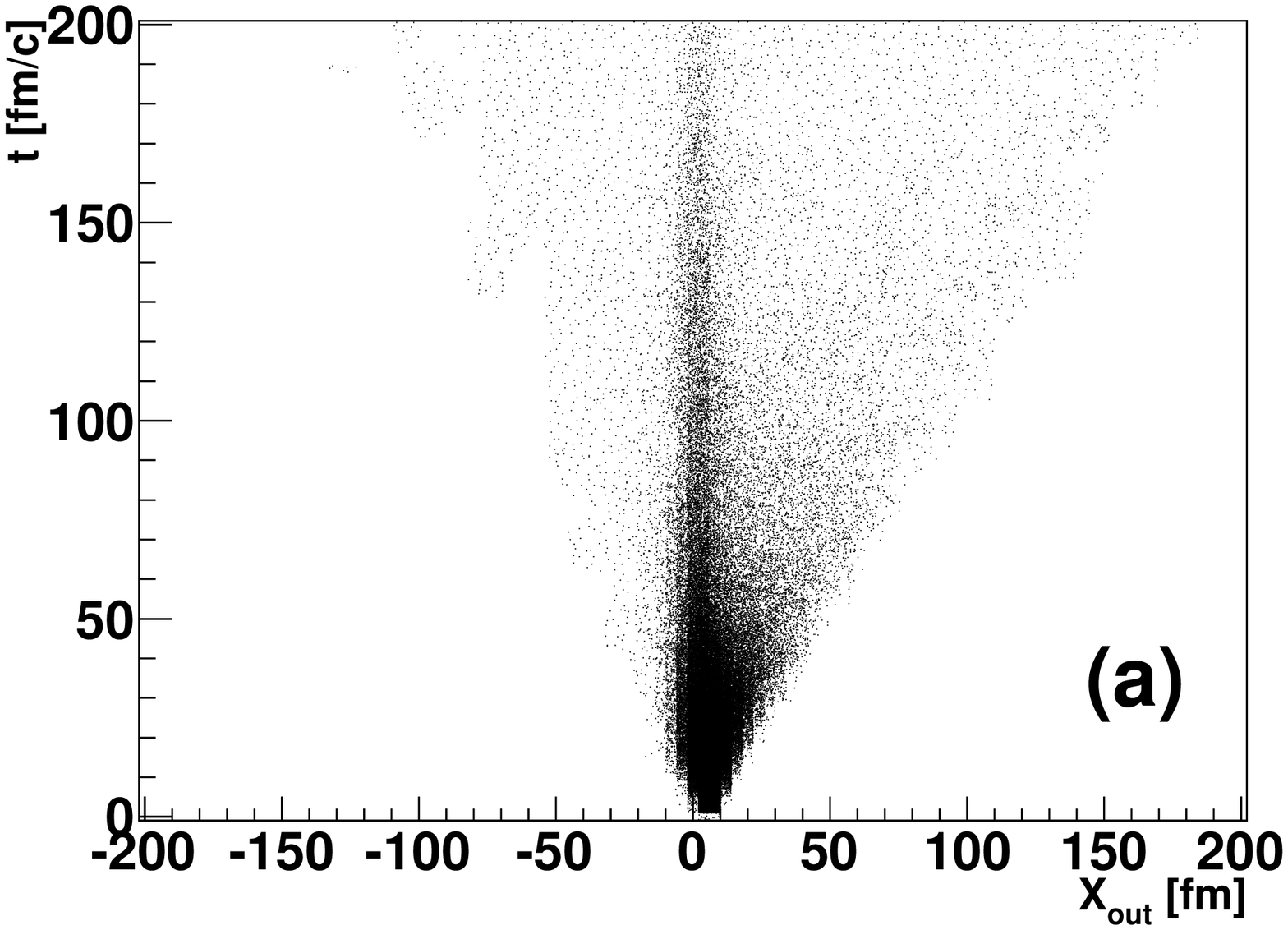}
\includegraphics[width=0.495\textwidth]{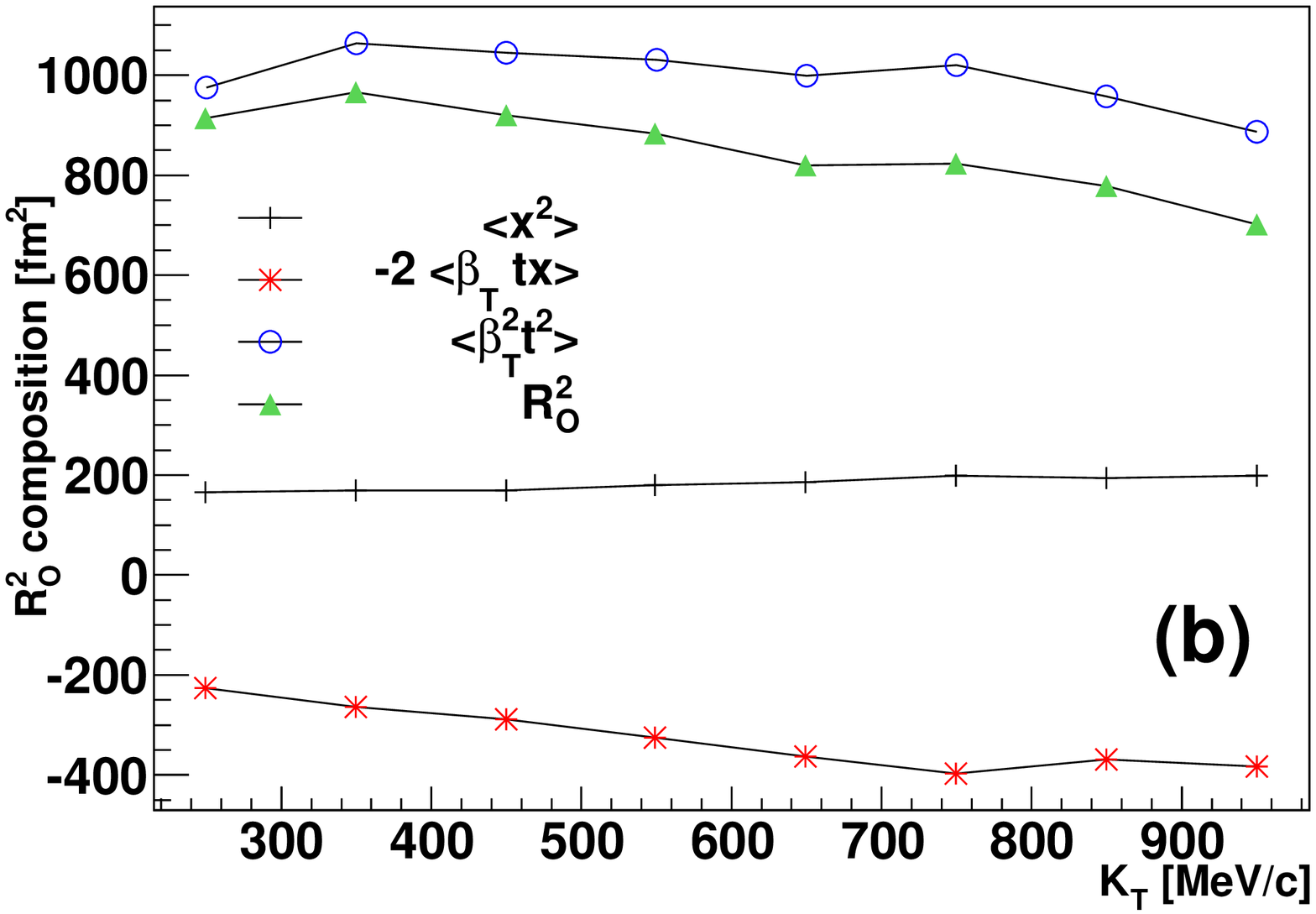}
\caption{(Color online) (a) the transverse position $x_{out}$ versus
emission time distribution of pions with cuts $|\eta|<0.8$, 200 $<
k_T < $ 300 MeV/c, $|q_i| < $ 100 MeV/c, and $t_{freezeout} <$ 199
fm/c. (b) the contribution of terms $<x^2>$ (line with crosses),
$-2<\beta_T t x>$ (line with asterisks), and $<\beta^2 t^2>$ (line
with open circles) in  Eq.~(\ref{eqn:RO}), to $R_O^2$ (line with
triangles) with same cuts as for (a). \label{fig2} }
\end{figure}

\begin{figure}
\includegraphics[angle=0,width=0.9\textwidth]{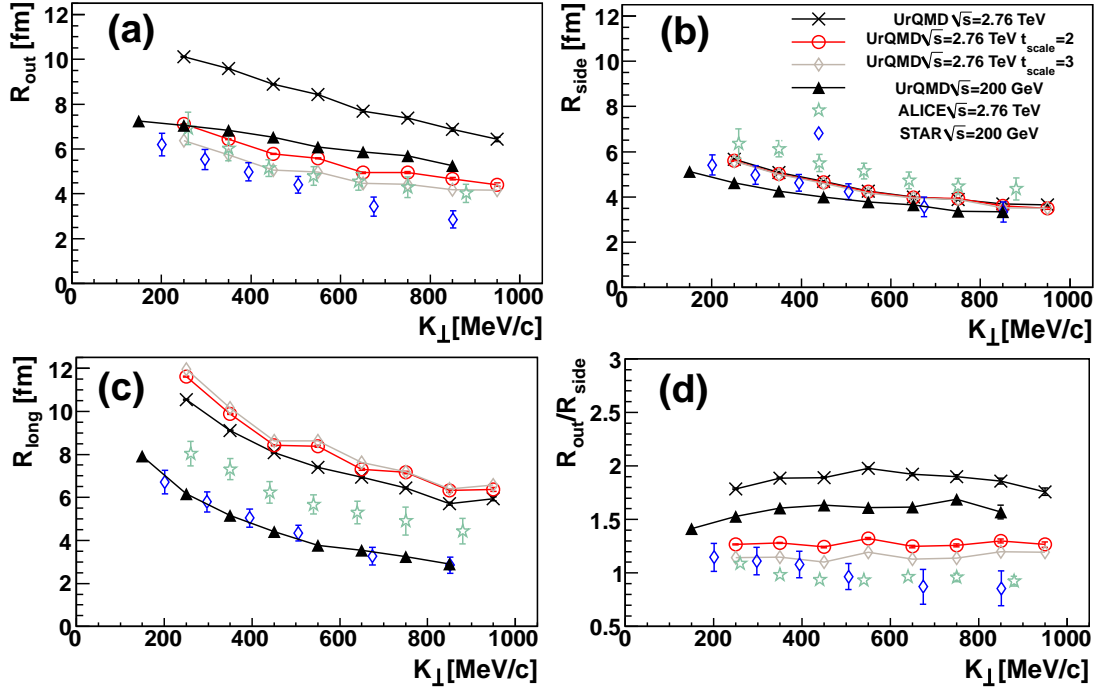}
\caption{ (Color online) $k_T$ dependence of pion HBT radii $R_O$
[panel (a)], $R_S$ [(b)], and $R_L$ [(c)], as well as the ratio
$R_O/R_S$ [(d)], for central ($\sigma/\sigma_T<5\%$) Pb-Pb
collisions at LHC energy $\sqrt{s_{NN}}=2.76$ TeV. For comparison,
parameters for central ($\sigma/\sigma_T<5\%$) Au-Au collisions at
RHIC energy $\sqrt{s_{NN}}=200$ GeV are also shown. Lines with
up-triangles and crosses are for model calculations while scattered
symbols are for experimental data of STAR/RHIC and ALICE/LHC
collaborations taken from \cite{Adams:2004yc,Aamodt:2011mr}. Lines
with circles and diamonds show results with an artificially
decreased emission duration by a factor of $t_{scale}=2$ and $3$,
separately, in the analysis of correlation function.} \label{fig3}
\end{figure}

Besides the non-Gaussian effect, the contribution of the correlation
between the emission time and position to the HBT radii, especially
in the outward direction, has been paid more attention in recent
years since it closely relates to the stiffness of the EoS of
nuclear matter especially at the early stage of the whole dynamic
process \cite{Lin:2002gc,Li:2007yd,Li:2008qm}. In Fig.\ \ref{fig2}
(a) we show the calculated emission time versus transverse position
$x_{out}$ of pions. The cuts $|\eta|<0.8$ and 200 $< k_T < 300$
MeV/c are adopted to have the same acceptance as for the extraction
of the HBT radii. At the same time, since the correlation function
and thus the HBT radii are mainly sensitive to pairs with small
momentum difference, a cut on the relative pair momentum $|q_i|<100$
MeV/c is applied as well. Further, in order to remove the
contribution of long-lived resonances, a cut on the freeze-out time
($t_{freezeout}<$ 199 fm/c) is used. It is found that, even in the
cascade calculation, there exists a visibly positive correlation
between the emission time and position. To further analyze the
importance of the $x-t$ correlation, we quantitatively calculate all
three terms in Eq.~(\ref{eqn:RO}) and show them in Fig.\ \ref{fig2}
(b) as a function of $k_T$. As a whole, although the magnitude of
the $x-t$ correlation term ($-2\langle \beta_T tx \rangle \approx$
-300 fm$^2$) is as big as that of the emission region term ($\langle
x^2 \rangle \approx$ 200 fm$^2$), the most important contribution to
$R_O$ comes from the emission duration term ($\langle \beta_T^2t^2
\rangle \approx$ 1000 fm$^2$). It implies that both a shorter
duration time and a stronger $x-t$ correlation lead to a smaller
$R_O$ value, which will be further discussed in Fig.\ \ref{fig3}
specialized for the result of HBT radii. Here, it is interesting to
see that the direct computation of $R_O$ leads to a value of
$\approx$ 30 fm which is larger than the value extracted from the
Gaussian fit to the correlation function by about a factor of three,
as was also observed in the AMPT model calculations for Au-Au
collisions at RHIC \cite{Lin:2002gc}.

Fig.\ \ref{fig3} shows the $k_T$ dependence of the HBT radii $R_O$,
$R_S$, and $R_L$, as well as the ratio $R_O/R_S$, extracted from the
Gaussian fit to the two-pion correlators. The UrQMD cascade
calculations for central Pb-Pb collisions at LHC energy
$\sqrt{s_{NN}}=2.76$ TeV (lines with crosses) and central Au-Au
collisions at RHIC energy $\sqrt{s_{NN}}=200$ GeV (lines with
up-triangles) are compared to corresponding experimental data by
ALICE/LHC (open stars) and STAR/RHIC (open diamonds). A strong
decrease of the three HBT-radii with $k_T$ is seen both in
experiments and in the UrQMD calculations for HICs. This implies a
substantial expansion of the source and is qualitatively captured by
the UrQMD dynamics. Following experimental results, the calculated
HBT radii for Pb-Pb at LHC are found to be larger than those for
Au-Au at RHIC. The largest increase exists in the longitudinal
direction, which is also seen by the experiments. Although the
comparison of the calculated HBT radii $R_L$ and $R_S$ with
corresponding data at RHIC is fairly well, it gets worse when going
to the higher LHC energy. At LHC the calculated $R_S$ values at all
$k_T$ are found to be slightly smaller than data, while $R_L$ and
$R_O$ values are larger than data. Together with large calculated
$R_O$, the emission time related quantity $R_O/R_S$ is found to be
markedly larger than the data.

From Eq.~(\ref{eqn:RO}) and Fig.~\ref{fig2} (b) it is clear that the
value of $R_O$ is strongly dependent on the emission duration of the
particles. To further investigate the contribution of the emission
duration to the HBT radii, we artificially decrease it by rescaling
the relative time $t$ to the ``effective source center time''
$\overline{t}$ ($=<t_i>$) by $t= t_i-\overline{t} \rightarrow
t'=(t_i-\overline{t})/t_{scale}$ in the calculation of the
correlation function at LHC energies. This effectively changes
Eq.~(\ref{eqn:RO}) to
\begin{equation}
 R_O'^2 = \langle (x-\beta_T t')^2 \rangle = \langle x^2 \rangle - 2 \frac{\langle \beta_T tx \rangle}{t_{scale}} + \frac{\langle \beta_T^2t^2\rangle}{t_{scale}^2}.
\end{equation}
The results for this calculation are presented as lines with circles
($t_{scale}=2$) and with diamonds ($t_{scale}=3$) in Fig.\
\ref{fig3}. The artificially decreased emission duration leads to
smaller $R_O$ values in all $k_T$ bins but leaves $R_S$ unchanged,
as expected. Overall it results in an improved agreement with the
data of $R_O/R_S$ ratio. From Fig.~\ref{fig3} it is also found that
$R_L$ is overestimated at LHC. Since $R_L$ is mainly related to the
lifetime of the source, it implies that this lifetime is also
overestimated by UrQMD. Other calculations in
\cite{Aamodt:2011mr,Graef:2012sh} show that UrQMD overestimates the
source lifetime by a factor of $\sim2-3$ when compared to LHC data.
The overestimation of both $R_O$ and $R_L$ can be attributed to the
known fact that the pressure in the early stage is not strong enough
in the cascade model calculations. A higher pressure would lead to a
more explosive expansion, a stronger phase-space correlation, and a
faster decoupling of the system, thus leading to smaller regions of
homogeneity. For more discussion we refer the reader to
\cite{Pratt:2008bc,Li:2007yd}. With the improved integrated
Boltzmann + hydrodynamics hybrid approach
\cite{Petersen:2008dd,Petersen:2011sb,Steinheimer:2010ib,Steinheimer:2011mp},
where various EoS of nuclear matter during the hydrodynamic
evolution may be treated consistently and a decoupling supplemented
by realistic 3d hypersurfaces we hope to get a satisfactory solution
in the near future.

\section{Summary}
To summarize, the two-pion HBT correlations (in the LCMS system) for
central Pb-Pb collisions at the LHC energy $\sqrt{s_{NN}}=2.76$ TeV
are calculated for the first time with the microscopic transport
model UrQMD. The non-Gaussian effect is seen especially in both
longitudinal and outward directions. Both the transverse momentum
$k_T$ dependence and the beam energy (from RHIC to LHC) dependence
of the HBT radii $R_O$, $R_S$, and $R_L$, extracted from a three
dimensional Gaussian fit to the correlator, exhibit qualitatively
the same behaviour as found in the experiments. However, the
calculated $R_O/R_S$ ratios at all $k_T$ bins are found to be larger
than in the data, both at RHIC \& LHC. We traced this finding back
to the explosive dynamics of the fireball at LHC which results in
both a shorter emission duration and a stronger time-space
correlation than modeled here.

\section*{Acknowledgments}
We would like to thank S. Pratt for providing the CRAB program. This
work was supported by the Helmholtz International Center for FAIR
within the framework of the LOEWE program launched by the State of
Hesse, GSI, and BMBF. Q.L. thanks the financial support by the key
project of the Ministry of Education (No. 209053), the NNSF (Nos.
10905021, 10979023), the Zhejiang Provincial NSF (No. Y6090210), and
the Qian-Jiang Talents Project of Zhejiang Province (No. 2010R10102)
of China. G.G. thanks the Helmholtz Research School for Quark Matter
Studies (H-QM) for support. Computational resources were provided by
the LOEWE-CSC.


\end{document}